\RequirePackage{fix-cm}
\documentclass[twocolumn]{svjour3}          

\pdfoutput=1

\smartqed  
\usepackage[usenames]{color}
\usepackage{graphicx}
\usepackage{graphics}
\usepackage{epsfig}
\usepackage{amssymb}
\usepackage{subfigure}
\usepackage{pifont}
 \journalname{Granular Matter}
\begin{document}

\title{Mechanical properties of inclined frictional granular layers}

\author{
A.P.F. Atman
\and
P. Claudin
\and
G. Combe
\and
G.H.B. Martins
}


\institute{
A.P.F. Atman \at
Departamento de F\'{\i}sica e Matem\'atica and National Institute of Science and Technology for Complex Systems, Centro Federal de Educa\c c\~ao Tecnol\'ogica de Minas Gerais, CEFET--MG, Av. Amazonas 7675, 30510-000, Belo Horizonte-MG, Brazil.\\
\email{atman@dppg.cefetmg.br}
\and
P. Claudin \at
Laboratoire de Physique et M\'ecanique des Milieux H\'et\'erog\`enes, (PMMH UMR 7636 CNRS -- ESPCI -- Univ. P. et M. Curie -- Univ. Paris Diderot), 10 rue Vauquelin, 75231 Paris Cedex 05, France.\\
\email{philippe.claudin@espci.fr}
\and
G. Combe \at
UJF-Grenoble 1, Grenoble-INP, CNRS UMR 5521, 3SR Lab., B.P. 53, 38041 Grenoble Cedex 09, France.\\
\email{gael.combe@ujf-grenoble.fr}
\and
G.H.B. Martins \at
Departamento de F\'{\i}sica e Matem\'atica, Centro Federal de Educa\c c\~ao Tecnol\'ogica de Minas Gerais, CEFET--MG, Av. Amazonas 7675, 30510-000, Belo Horizonte-MG, Brazil.
}

\date{Received: date / Accepted: date}

\maketitle

\begin{abstract}
We investigate the mechanical properties of inclined frictional granular layers prepared with different protocols by means of DEM numerical simulations. We perform an orthotropic elastic analysis of the stress response to a localized overload at the layer surface for several substrate tilt angles. The distance to the unjamming transition is controlled by the tilt angle $\alpha$ with respect to the critical angle $\alpha_c$. We find that the shear modulus of the system decreases with $\alpha$, but tends to a finite value as $\alpha \to \alpha_c$. We also study the behaviour of various microscopic quantities with $\alpha$, and show in particular the evolution of the contact orientation with respect to the orthotropic axes and that of the distribution of the friction mobilisation at contact.
\end{abstract}

\section{Introduction}

The nature of the jamming transition in granular systems has been investigated during the last decade, see recent reviews \cite{vH10,LNvSW10}. Many studies have focused on frictionless discs or spheres, typically controlled in volume fraction $\phi$ or in pressure $P$ \cite{oHLLN02,oHSLN03,MSLB07}, showing that the jamming transition is critical (scaling exponents, diverging length scale) \cite{oHLLN02,WNW05,EvHvS09} and related to isostaticity \cite{R00,TW99,M01,oHLLN02,AR07}. As the system loses its mechanical rigidity at the transition, its shear modulus $G$ is found to vanish as a power law with respect to the distance to jamming $\phi-\phi_c$, where $\phi_c$ is the critical volume fraction. The properties of frictional granular packings have also been investigated, see e.g. \cite{SEGHL02}, but, in this context of elastic properties close to jamming, most of the studies have considered homogeneous systems under isotropic pressure \cite{ZM05,AR07,SvHESvS07,HvHvS10,HSvSvH10,S10,BDBB10}. In the frictional case, the Liu-Nagel jamming concept \cite{LN98,LN10} must be revised \cite{BZCB11}. In particular, jamming and isostatic points do not coincide any more \cite{vH10}, and one thus can expect a finite shear modulus at the transition.

In this paper, we consider static layers of frictional grains under gravity, by means of two-dimensional discrete element simulations (standard Molecular Dynamics \cite{DEM}), and investigate their mechanical properties through the analysis of their stress response to a localized overload $\vec F_0$ at the layer surface, a technique particularly developed by and dear to R.P. Behringer, see e.g. \cite{GHLBRVCL01,ABGRCCBC05}. Expanding the work published in \cite{ACCM13}, we present here the detailed analysis of layers prepared with three different protocols. The outline is as follows. We first describe the numerical system, its preparation and the computation of the stress response. In the next section, we present an orthotropic elastic analysis of the stress profiles, and detail the fitting procedure. Then, a section is devoted to the measure and the interpretation of the microscopic data. Finally, conclusions and perspectives are drawn.

\section{Numerical simulations}

\subsection{Numerical model and set-up}
\begin{figure}[t]
\includegraphics[width=\linewidth]{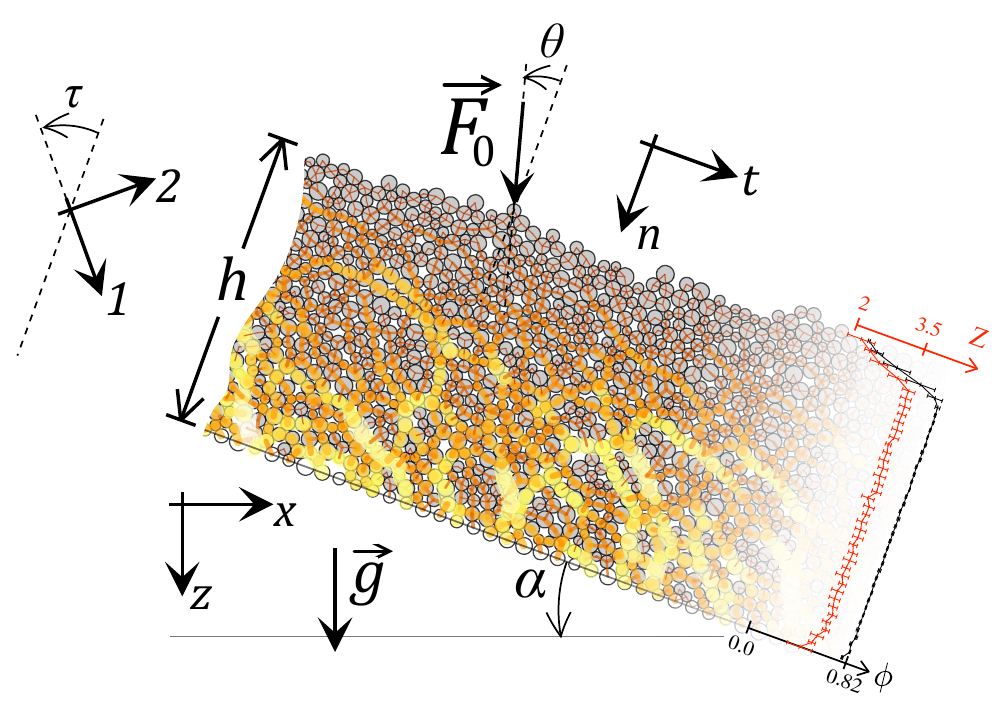}
\caption{(color online) System set-up and notations. $x$ is the horizontal axis. $z$ is the vertical one, along which acts gravity $\vec g$. The granular layer (here GG preparation), of average thickness $h$, is inclined at an angle $\alpha$ with respect to horizontal. $t$ and $n$ are the axis respectively tangential and normal to the layer.  A localized force ${\vec F}_0$, which makes an angle $\theta$ with respect to $n$, is applied on a grain close to the surface of the layer. The stress responses $\sigma_{nn}$ and $\sigma_{tn}$ to this overload are measured at the bottom of the layer (fixed grains in white). Axis $(1,2)$, making an angle $\tau$ with respect to $(n,t)$, are those of the orthotropic elastic analysis. Black line: volume fraction profile $\phi(n)$. Red line: coordination number profile $Z(n)$. These are for the GG preparation. Orangish colors on grains: force chains.}
\label{SetUp}
\end{figure}

The numerical model is that described in \cite{ABGRCCBC05,GACCG06}, with $N=3600$ polydisperse frictional discs coupled, when overlapping, by normal and tangential linear springs, tangential forces being limited by the Coulomb condition with a friction coefficient $\mu=0.5$. The typical thickness of the layer is $h \simeq 23$ grain diameters, i.e. a system aspect ratio around $1/6$. The layers are prepared at a \emph{fixed} angle $\alpha$ with respect to the horizontal (see Fig.~\ref{SetUp} for notations), and unjamming is approached as $\alpha$ is close to $\alpha_c$, the critical value above which static layers cannot be equilibrated at that angle and always flow. Note that this unjamming point $\alpha_c$ is close in spirit to the situation of a jammed solid sheared up to its yield-stress~\cite{HB09}. It is also close, but different, to progressively tilted granular layers, which eventually loose their mechanical stability, see e.g., \cite{SVR02,HBDvS10}.

In our simulations, the volume fraction in the layer is fairly uniform all through the layer depth and roughly independent on the inclination angle. The control parameter for the jamming/unjamming transition is then the sole angle $\alpha$. This situation is therefore qualitatively different to the homogeneous configurations submitted to isotropic pressure cited above, and is effectively closer to an experimental set-up. No external pressure applied to the topmost layer of particles, i.e. the pressure in the system is due solely to the gravitational force acting on the particles themselves.

\subsection{Three preparation protocols}
Three different system preparations have been carried out: a grain-by-grain (GG), a rain-like (RL) and an avalanched (AV) deposition of the particles on a rough substrate consisting of fixed but size-distributed particles, inclined at the desired angle $\alpha$. In the GG protocol, grains are added to the layer one after the other, with no initial velocity, at random $t$-positions and in contact with those already deposited. The time lag between two successive drops is sufficiently large to ensure the relaxation of the system before the next deposit. As for the RL preparation, all $N$ grains are initially put at regular `flying' positions above the bed, with no contact between the particles and no velocity. Then gravity is switched on, and they all fall down like a rain. Finally, for the AV preparation, we start from an initial steady and homogeneous flow running at a large inclination, then abruptly set the angle to the desired value of $\alpha$ and reduce the kinetic energy of the whole system. The layer is prepared when all grains have eventually reached static equilibrium (see \cite{ABGRCCBC05} for more details).

Above a certain inclination $\alpha_c$, these preparation procedures do not converge towards a static layer -- the grains do not stop moving. The `solid-liquid' transition occurs rather abruptly, over a typical inclination range $\Delta \alpha \simeq 0.5^\circ$ where only part of the simulations converge. This allows for a value of this critical angle defined at this precision. For both GG and AV preparations, we get $\alpha_c \simeq 20.8^\circ$. We have not studied systematically enough the RL preparation for inclinations around $20^\circ$ to determine its critical angle with a good precision. However, we expect RL and AV data to be very similar close to $\alpha_c$ as in both cases the grains flow down the slope over long distances -- typically several times the system size -- before stoping, so that the initial configuration is effectively forgotten.

\begin{figure*}[t]
\includegraphics[width=\linewidth]{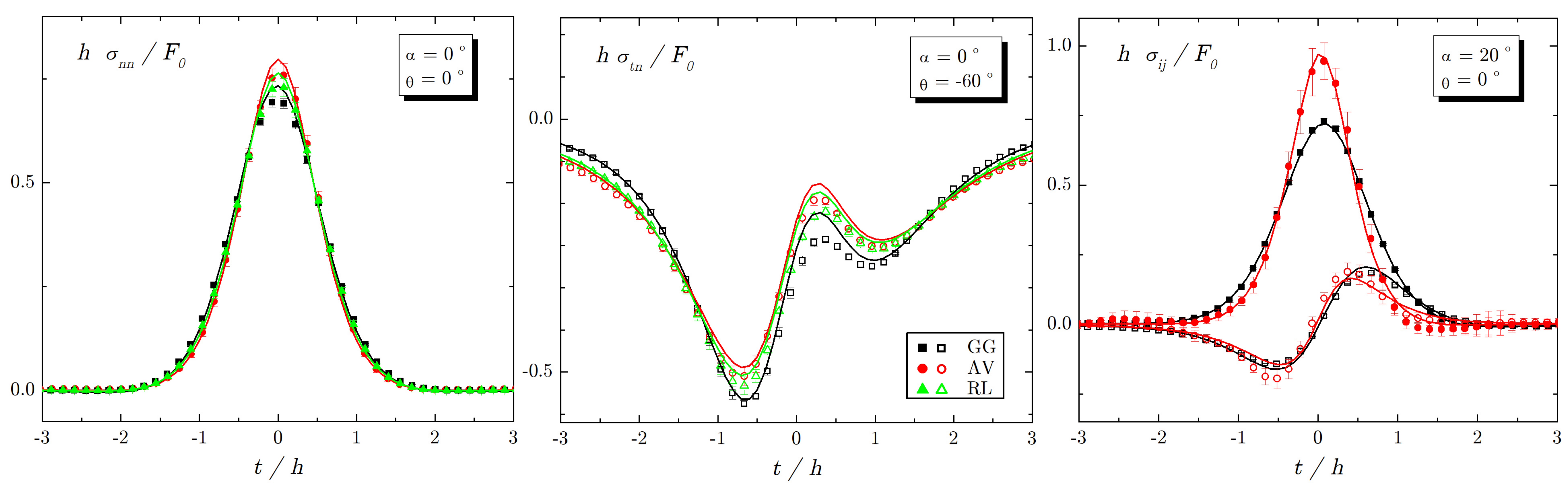}
\caption{(color online) Stress profiles for the different preparations. The layer inclination $\alpha$ and the overload angle $\theta$ are indicated in legend for each panel. Symbols: numerical data (filled symbols: $\sigma_{nn}$; empty symbols: $\sigma_{tn}$; color code: see legend). Lines: elastic fits (see table~\ref{FittingParam} for the corresponding values of the fitting parameters).}
\label{StressProfiles}
\end{figure*}

\begin{table}[t]
\begin{center}
\begin{tabular}{|c|c|c|c|c|c|}
\hline
$\alpha$	& prep.	& $G/E_1$	& $E_2/E_1$	& $\nu_{21}$     & $\tau$ \\
\hline
$0^\circ$	& GG	& 0.403		& 0.80		& 0.20	          & $93^\circ$ \\
		& RL		& 0.303		& 0.69		& 0.23	          & $93^\circ$ \\
		& AV		& 0.275		& 0.71		& 0.26	          & $91^\circ$ \\
\hline
$20^\circ$	& GG	& 0.262		& 0.49		& 0.17	          & $66^\circ$ \\
		& AV		& 0.248		& 0.93		& 0.27	          & $94^\circ$ \\
\hline
\end{tabular}
\end{center}
\caption{Values of the elastic parameters corresponding to the fits displayed in Fig.~\ref{StressProfiles}.}
\label{FittingParam}
\end{table}

These three preparations mainly differ in their contact orientation (see Fig.~\ref{MicroData}). Their volume fractions does not vary much from $\alpha=0^\circ$ to $\alpha_c$. Typical values are $\phi \simeq 0.82$ for GG and $\phi \simeq 0.81$ for RL and AV. These are slightly larger than -- or similar to --  the critical value, estimated in our system at $\phi_c \simeq 0.81$ \cite{dCEPRC05,S10,OH11}.

\subsection{Stress response profiles}
Once a layer is deposited, stabilized in an equilibrium state, an additional force $\vec F_0$ is applied on a grain close to the free surface, and a new equilibrium state is reached. Taking the difference between the states after and before the overload, one can compute the contact forces in response to $\vec F_0$. Introducing a coarse graining length $w$, the corresponding stress response can be determined. Taking $w$ of the order of few mean grain diameters (here $w =  6 \left < d \right >$) as well as an ensemble averaging of the data (here, for each tilt angle $\alpha$, we average over $120$--$150$ independent force loads, distributed on typically $10$ layers in total), make the stress profiles quantitatively comparable to a continuum theory \cite{GACCG06}, such as elasticity, as discussed below. The amplitude of the overload was kept constant for all simulations: $F_0 =  1.0 \left < m \right > g$, where $\left < m \right >$ is the average mass of the grains. This value is  sufficiently small to ensure a linear \cite{ACCGG09,ACC09} and reversible response of the system for all values of $\alpha$, including close to $\alpha_c$.

Some examples of stress bottom profiles $\sigma_{nn}(t)$ and $\sigma_{tn}(t)$ are displayed in Fig.~\ref{StressProfiles} for different values of the inclination $\alpha$ and of the angle $\theta$ that the overload force makes with the normal direction (see Fig.~\ref{SetUp}). Note that, as we deal with linear elasticity, the stresses can be rescaled by $F_0/h$. The normal stress data $\sigma_{nn}$ show classical bell-shaped profiles, which do not differ much for all three preparations when the layer is horizontal ($\alpha=0$) and the overload vertical ($\theta=0$), see panel (a). However, on can distinguish between the preparations, especially GG from the two others, looking at the shear stress profiles $\sigma_{tn}$ in response to a non-normal overload force ($\theta=-60^\circ$), see panel (b). The difference between GG and AV profiles is enhanced for the data at an inclination close to $\alpha_c$,  see panel (c).

\begin{figure*}[t]
\centerline{\includegraphics[width=0.9\linewidth]{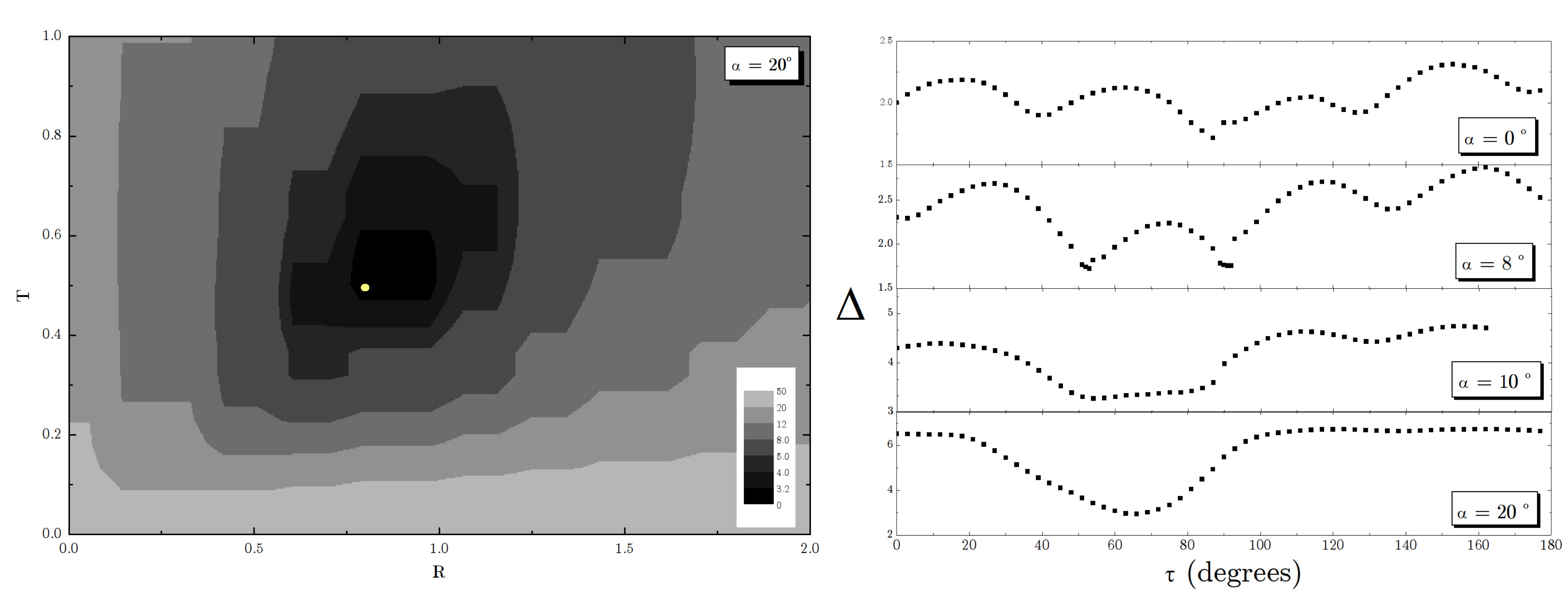}}
\caption{Fitting technique. (a) Contour plot, in the $(R,T)$ parameter plane, of the normalized difference $\Delta$ (Eq.~\ref{DefDelta}) between the numerical data and the elastic prediction. The other parameters are $\nu_{21}=0.15$ and $\tau=66^\circ$. The layer inclination is $\alpha=20^\circ$. White bullet: location of the best fit. (b) Difference $\Delta$ as a function of the orthotropic angle $\tau$ for four values of $\alpha$ (see legends). These are GG data. For each of these points, all other parameters are also set to their best fitting values.}
\label{FittingTechnique}
\end{figure*}

\section{Orthotropic elastic analysis}

Experimental and numerical works have shown that the linear stress response of granular systems to a point force is well described by (possibly anisotropic) elasticity \cite{SRCCL01,ABGRCCBC05,GG02,LTWB04,GG05,ABGRCBC05,GWM06}. In this section, we introduce the framework of orthotropic elasticity, with which numerical response profiles such as those displayed in Fig.~\ref{StressProfiles} can be fitted. The details of the computation of elastic response are available in Appendix~\ref{orthotropic}.

\subsection{Orthotropic elasticity}
Orthotropic elasticity is characterized by a stiff axis (here labelled $1$) and a soft one (labelled $2$), associated to two Young moduli $E_1$ and $E_2<E_1$, and to two Poisson coefficients $\nu_{12}$ and $\nu_{21}$ (note that, for symmetry reasons, $\nu_{12}/E_1 = \nu_{21}/E_2$). There is also a shear modulus $G$ involved in the corresponding relation between stress and strain tensor components (Eq.~\ref{Wdagger}). A last parameter of this modeling is the angle $\tau$ that the axes $(1,2)$ make with $(n,t)$ (see Fig.~\ref{SetUp}).

Orthotropic stress responses to a point force ${\vec F}_0$ have been analytically computed in \cite{OBCS03} for a semi-infinite medium ($h \to \infty$). For a given $\tau$, they only depend on two combinations of the elastic parameters, noted $R$ and $T$, (Eq.~\ref{defTandR}). For an elastic slab of finite layer thickness $h$, a semi-analytical integration, following the computation performed in \cite{SRCCL01} for isotropic elasticity, must be done (see Appendix~\ref{orthotropic}). Rough bottom boundary conditions (zero displacement) are imposed. Besides the coefficients $R$ and $T$, these bottom conditions involve a Poisson coefficient, say $\nu_{21}$, so that, in total, five dimensionless numbers ($\tau$, $R$, $T$, $\nu_{21}$ and $\theta$) must be specified to produce the normalized bottom stress responses $\sigma_{ij} h/F_0$ as functions of the reduced tangential coordinate $t/h$. 

\subsection{Fitting numerical data}
The idea is to fit the elastic response profiles to the numerical data, in order to extract the effective elastic parameters of the layer. For a given inclination $\alpha$, the four numbers $\tau$, $R$, $T$ and $\nu_{21}$ must be adjusted to reproduce at the same time the profiles measured for all three stress components $\sigma_{nn}$, $\sigma_{tn}$ and $\sigma_{tt}$, and for all overload angles $\theta$. This is achieved by minimizing the RMS difference
\begin{equation}
\Delta = \sqrt{\frac{1}{N_p} \,\, \sum_{\{i,j\}, \, \theta} \,\, \sum_{k=1}^{N_p} \left ( \frac{\left . \sigma_{ij}^k \right |_{\rm num} - \left . \sigma_{ij}^k \right |_{\rm elas}}{\delta \sigma_{ij}^k} \right )^2},
\label{DefDelta}
\end{equation}
where $N_p$ is the number of data points in the profiles, and $\delta \sigma_{ij}$ is the standard deviation around the mean stress computed from the ensemble averaging.

An example of a contour plot of $\Delta$ in the ($R, T$) plane, for given $\tau$ and $\nu_{21}$, is shown in Fig.~\ref{FittingTechnique}a. There is a clear deepest point, which corresponds to the best fit. In Fig.~\ref{FittingTechnique}b, we display $\Delta$ as a function of the orthotropic angle $\tau$, each point of these curves corresponding to the best fitting $R$,$T$ and $\nu_{21}$. These curves have been computed for the GG data at different inclination angles. It shows how the minimum, corresponding to the best fitting $\tau$, changes rather abruptly from $\simeq 90^\circ$ to $\simeq 60^\circ$ around $\alpha \simeq 9^\circ$ (see also next section and Fig.~\ref{MicroData}c).

\begin{figure*}
\includegraphics[width=0.48\linewidth]{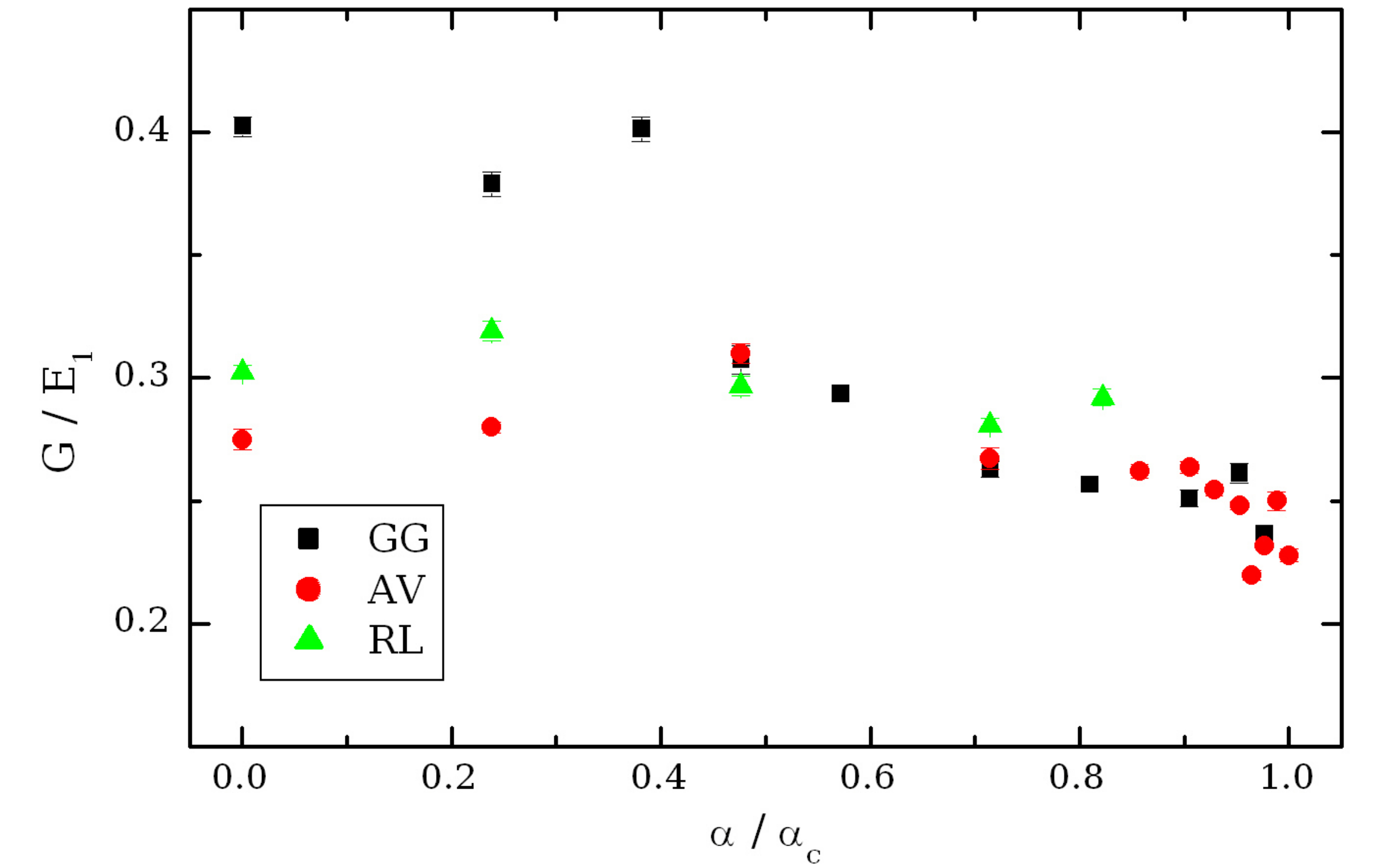}
\hfill
\includegraphics[width=0.48\linewidth]{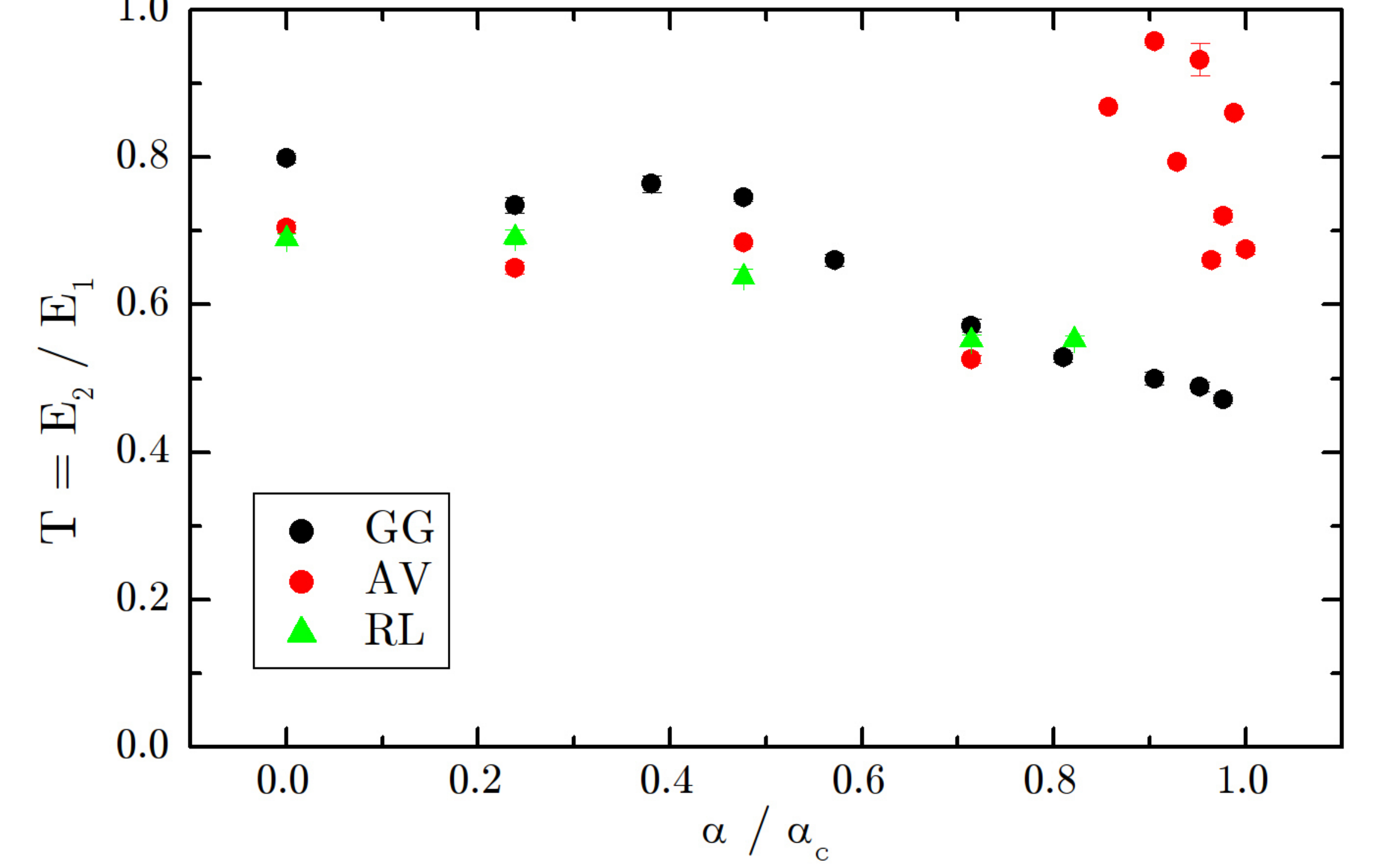}
\caption{(color online) Shear and Young moduli ratios $G/E_1$ (a) and $E_2/E_1$ (b) as functions of $\alpha/\alpha_c$. These data include all three preparations GG, RL and AV, see legend.}
\label{ShearYoungModuli}
\end{figure*}

Some of these fits are displayed in Fig.~\ref{StressProfiles}, for various angles $\alpha$ and $\theta$, and for the different preparations. The overall agreement between the elastic predictions and the numerical data is quantitatively good. In Fig.~\ref{ShearYoungModuli}, we show the elastic modulus ratios $G/E_1$ and $E_2/E_1$ extracted from these fits, as function of the inclination. $G/E_1$ decreases with $\alpha$ but does not vanish close to the critical angle, in agreement with the observation that frictional granular systems remain hyperstatic at the unjamming transition \cite{AR07,SvHESvS07,HvHvS10}. Such a discontinuous behaviour at the transition has also been seen in simulations by Otsuki and Hayakawa \cite{OH11} investigating the rheology of sheared frictional grains close to jamming, and in experimentally created shear-jammed states reported in \cite{BZCB11}. The sudden drop of $G/E_1$ around $\alpha \simeq 9^\circ$ is associated with the change of the orthotropic directions mentioned above. The behaviour of $E_2/E_1$ also present an overall decrease with $\alpha$, except for the AV data close to $\alpha_c$. The complete interpretation of this behavior of the AV data is not entirely clear, but it is clearly related to an increase of friction mobilization at the contacts (see Figs.~\ref{MicroData} and \ref{ContactsGlissants} and discussion below).

\section{Microscopic variables}

\begin{figure*}[t]
\includegraphics[width=\linewidth]{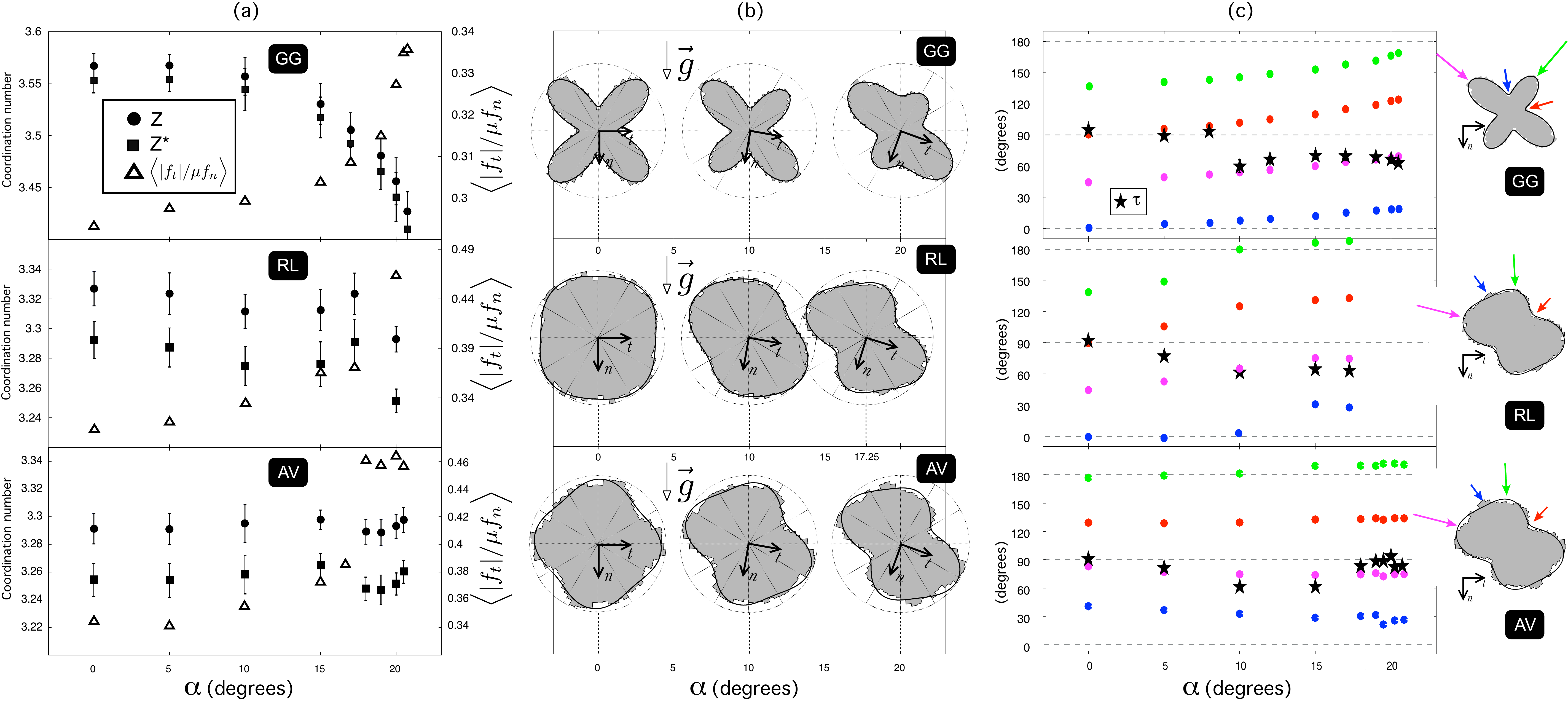}
\caption{(color online) Microscopic data for the three preparation protocols GG (top), RL (middle) and AV (bottom). (a) Coordination number $Z$ (\ding{108}) and modified (`rattlers' removed) coordination number $Z^*$ ($\blacksquare$) as functions of the inclination of the layer $\alpha$. Right $y$-axis: relative importance of the average friction mobilisation at contact ($\bigtriangleup$). (b) Contact angle polar distributions at three inclination angles $\alpha$. Solid black line: fourth-order Fourier fit. Gravity is vertical (black arrow). (c) Fitted orthotropic elastic angle $\tau$ as a function of $\alpha$ ($\bigstar$). The four characteristic angles of the contact angle distribution, computed with respect to the direction $n$, are also shown -- these angles corresponds to the directions of the lobes, and those in between the lobes, see sketch and corresponding coloured arrows in legend.}
\label{MicroData}
\end{figure*}

In addition to the above global mechanical properties of the system, we have studied the evolution of various microscopic quantities with $\alpha$. The first one of interest is the coordination number $Z$, i.e. the average number of contacts per grain, here computed in the bulk of the layer, where it is fairly uniform -- it obviously drops down close to the surface. $Z$ monotonously decreases with $\alpha$ for the GG preparation, while it stays approximately constant for RL and AV data (Fig.~\ref{MicroData}a). In all cases, it stays always far from the isostatic value $Z_{\rm iso}=3$ (for frictional grains in 2D). Grains of the bulk that only carry their own weight do not contribute much to the global stability of the contact network. As for so-called rattlers in gravity-free packings (see \cite{bookDEM}, chap. 6), these grains can be removed from the contact counting, leading to a modified coordination number of the layer $Z^*$ (see Fig.~\ref{MicroData}a). However, we find that their number is roughly independent of $\alpha$.

\begin{figure*}[t]
\includegraphics[width=0.3\linewidth]{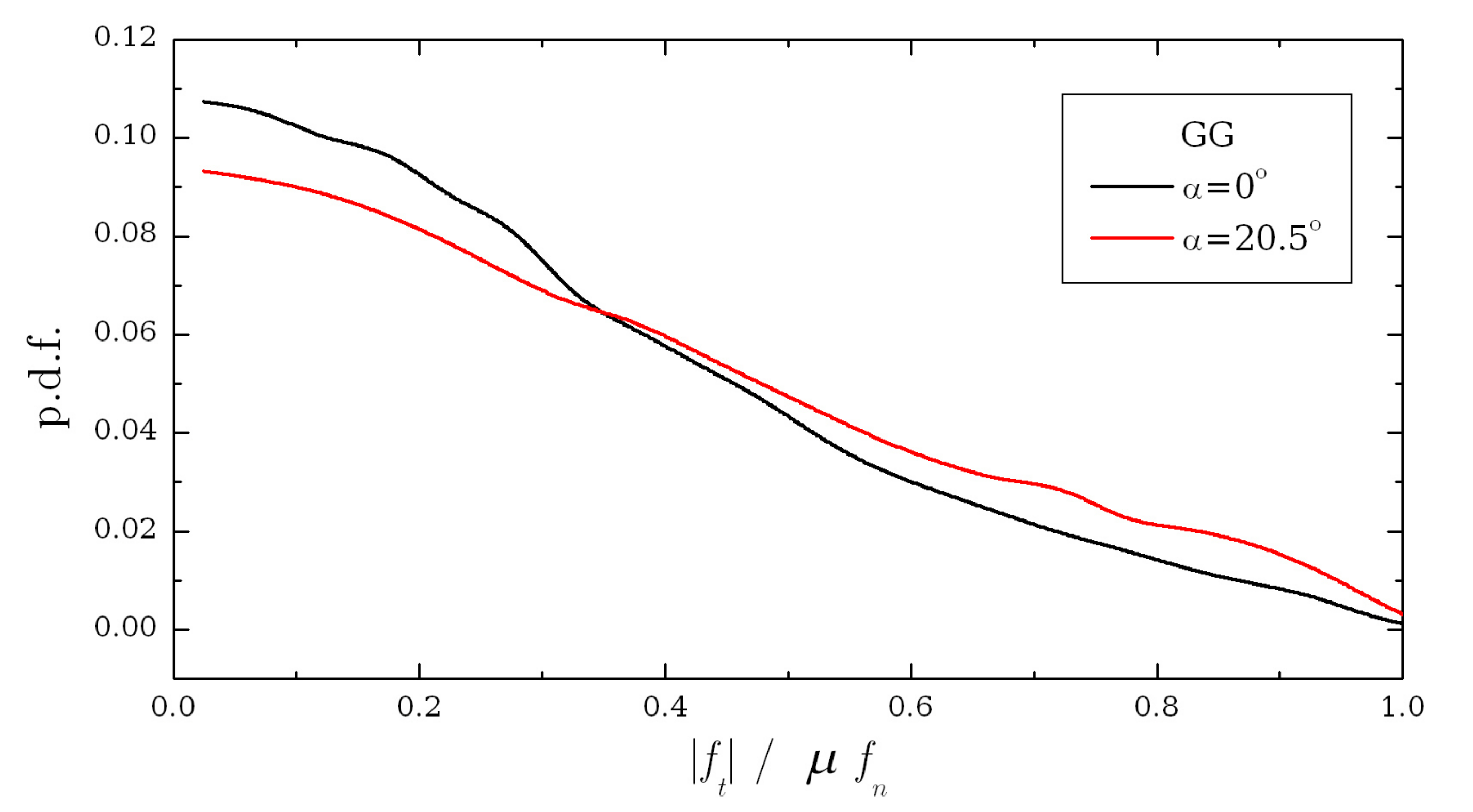}
\hfill
\includegraphics[width=0.3\linewidth]{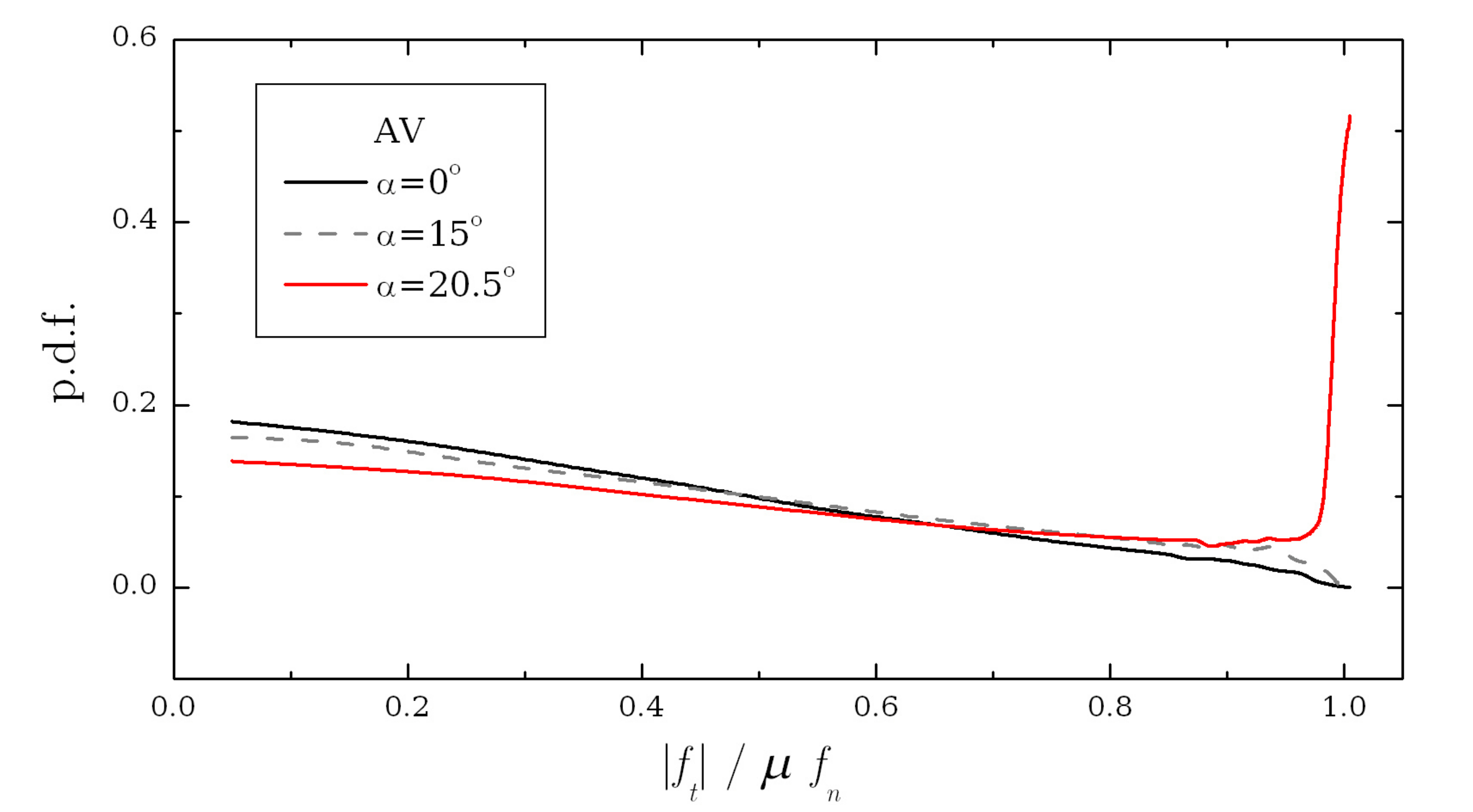}
\hfill
\includegraphics[width=0.3\linewidth]{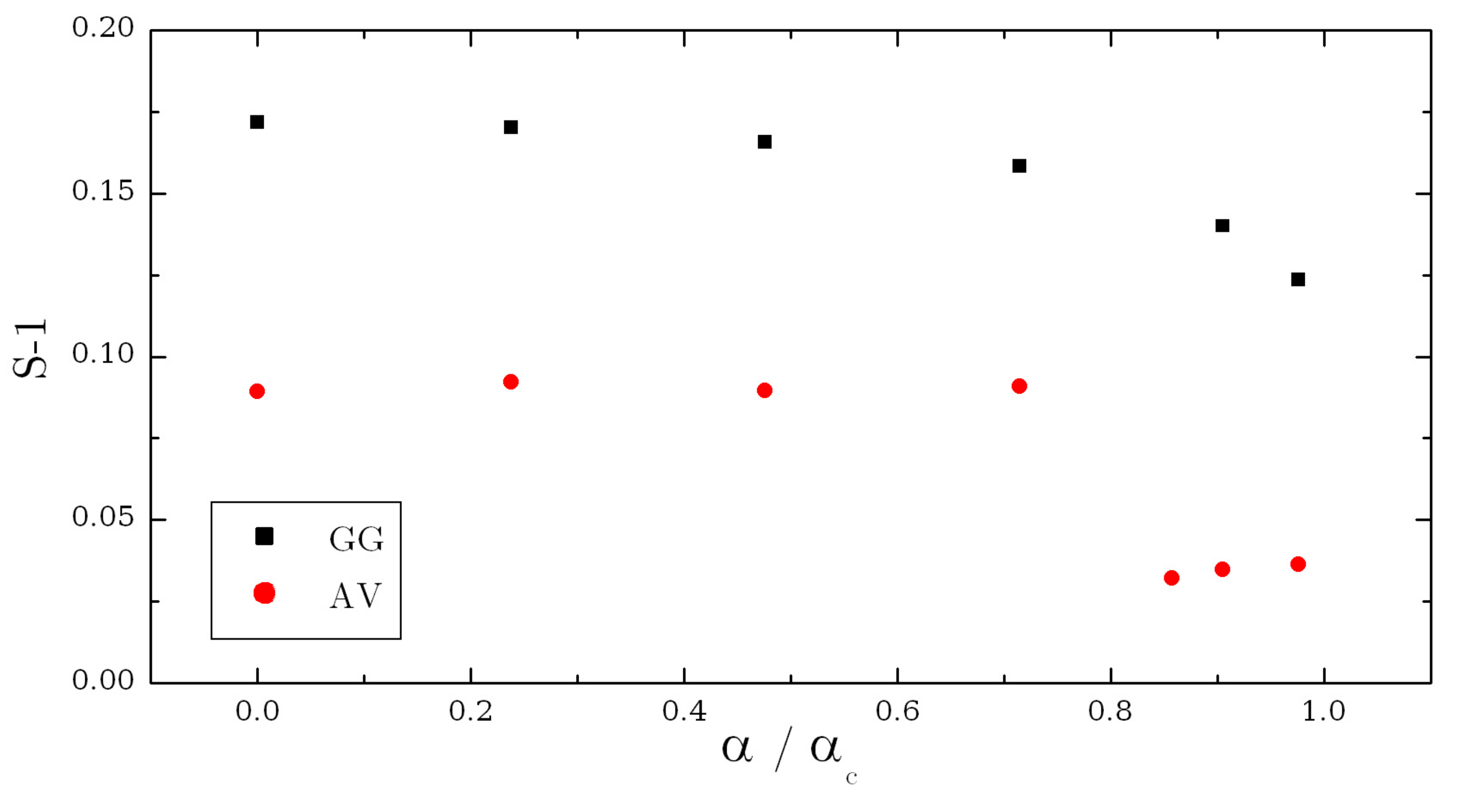}
\centerline{\includegraphics[width=0.7\linewidth]{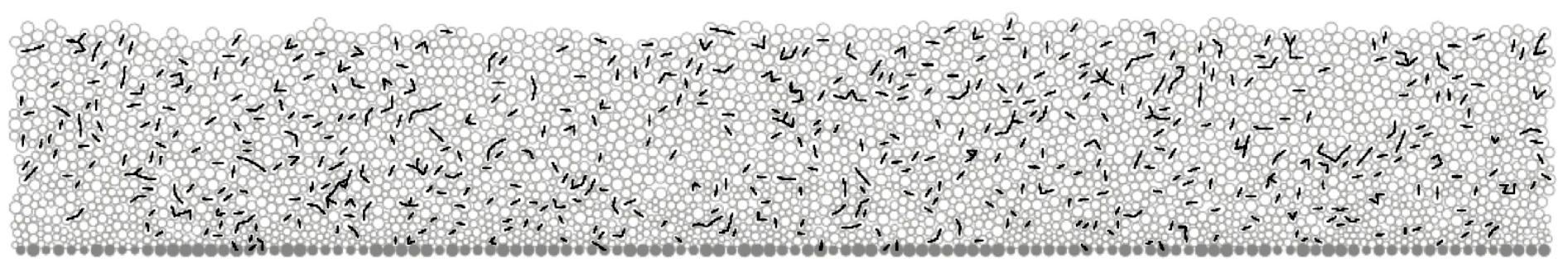}}
\caption{(color online) Probability distribution function of the friction mobilisation at contact $|f_t|/(\mu f_n)$ for the GG (a) and the AV (b) preparations. The distributions for several values of $\alpha$ are displayed. For the AV preparation, the distribution at $\alpha=18^\circ$ is not shown but is identical to that at $\alpha=20.5^\circ$. (c) Redundancy factor, as defined in \cite{K10}, as a function of $\alpha/\alpha_c$. (d) Spatial distribution of quasi-sliding contacts (bold dashes) in an AV-layer at $\alpha=20.5^\circ$.}
\label{ContactsGlissants}
\end{figure*}

We have also studied the friction mobilisation at the contact level. In the MD simulations, the number of contacts with a ratio of the tangential force $f_t$ to the normal force $f_n$ strictly equal to the microscopic friction $\mu$ is zero when static equilibrium is reached. However, some of them are effectively close to the Coulomb criterion. We have first computed the average $\langle \frac{|f_t|}{\mu f_n} \rangle$. This quantity, displayed in Fig.~\ref{MicroData}a, increases as $\alpha \to \alpha_c$ for all three preparations, but its overall variation is weaker for the GG data (see right $y$-scales), as could be expected. More precisely, we also display in Fig.~\ref{ContactsGlissants}a,b the probability distribution function of  of the friction mobilisation at contact for the two preparations GG and AV, and for several inclinations. For the GG preparation, the distribution is only slightly skewed towards larger values of $|f_t|/\mu f_n$ when $\alpha$ in increased, but nothing particular happens close to $|f_t|/\mu f_n=1$. For the AV preparation, however, a peak close to $|f_t|/\mu f_n=1$ appears for $\alpha \gtrsim 18^\circ$, corresponding to quasi-sliding contacts. Fig.~\ref{ContactsGlissants}d shows that they are uniformly distributed all through the layer depth. Following \cite{SvHvS07,HvHvS10,K10}, we have computed the redundancy factor $S$, i.e. the ratio of the total number of force degrees of freedom at contacts over the number of equilibrium equations, taking into account these quasi-sliding contacts: $S=(2 n_c - n_s)/(3N)$, where $n_c$ is the total number of contacts and $n_s$ is the number of quasi-sliding contacts -- recall the system is two-dimensional. We see that $S$ decreases with $\alpha$ (see Fig.~\ref{ContactsGlissants}c), and, for the AV preparation, approaches $1$ (the isostatic value), though remaining above this value at $\alpha_c$.

Finally, we have studied contact angle distributions. Three of these distributions are represented as polar diagrams for $\alpha=0$, $10$ and $20$ (or $17.25$ for RL) degrees in Fig.~\ref{MicroData}b. Let us first comment the GG data. The four strongly pronounced lobes are typical of this preparation \cite{bookDEM} (chap. 6). The vertical and horizontal directions are always in between these lobes. When the layer is horizontal ( $\alpha=0^\circ$), the orthotropic stiff and soft directions are also found to be (almost) along the horizontal and vertical axis respectively. Note that the fitting procedure effectively gives here $\tau=93^\circ$ in this case, while $\tau=90^\circ$ (or $0^\circ$) would have been expected for symmetry reasons. This effectively indicates the typical precision we have on the measure of this orthotropic angle. Close to the critical slope, however, the orthotropic orientations are close to those of the lobes, the stiff one being in the direction of the slope. As evidenced in Fig.\ref{MicroData}c, the transition between these two microscopic configurations occurs around $\alpha \simeq 9^\circ$, i.e. well below $\alpha_c$, in correspondence with the drop of $G/E_1$ between $8$ and $10^\circ$ (see Fig.~\ref{ShearYoungModuli}). The polar distributions computed with RL and AV data are more isotropic than in the GG case (Fig.~\ref{MicroData}b). However, although the lobes are less pronounced, the overall behaviour of the RL data is similar to the GG ones. In the AV case, the orthotropic direction roughly follows that of the lobes over the all range of inclination.

\section{Conclusions}

To sum up, we have simulated 2D frictional and polydisperse granular layers under gravity inclined at an angle $\alpha$, and investigated their mechanical and microscopic properties when the unjamming transition is approached. This work tells us what to expect in real experiments, i.e. a layer that becomes elastically softer as $\alpha \to \alpha_c$, as e.g. inferred from acoustic experiments on a granular packing in the vicinity of the transition \cite{BAC08}. More precisely, the shear modulus $G$ and the stiff Young modulus $E_1$ both decrease with respect to the soft modulus $E_2$, but not to the point at which the system would loose its rigidity before avalanching. In particular, as evidenced by the comparison of the curves in figures~\ref{ShearYoungModuli} and \ref{MicroData}a, the shear modulus is not found to be a linear function of $Z-Z_{\rm iso}$ (or $Z^*-Z_{\rm iso}$), in contrast with the finding of \cite{SvHESvS07} on homogeneous frictional systems, close to isostaticity. In fact, in agreement with the analysis of \cite{HBDvS10}, the idea that the whole granular layer reaches the isostatic limit at the critical angle $\alpha_c$ is too simple because it ignores the anisotropy and inhomogeneity of the packing induced by the preparation and the gravity field. Interestingly, in the simple shear geometry considered in \cite{K10}, the redundancy factor $S$ does tend to $1$ when the critical state is reached, but here remains (slightly) above this value for the avalanched layers, even though some (quasi) sliding contacts appear.

As for perspectives, similarly to what we did for the GG layers in \cite{ACCM13}, one should compute the vibration modes for the AV layers, taking into account the presence of these quasi-sliding contacts. Also, it could be interesting to use granular simulations with a rolling resistance \cite{ETR08} in order to explore a wider range of $\phi$, $Z$ and $\alpha$.

\begin{acknowledgements}
We thank I. Cota Carvalho, R. Mari and M. Wyart for fruitful discussions. This work is part of the ANR JamVibe, project \# 0430 01. A.P.F. Atman has been partially supported by the exchange program `Science in Paris 2010' (Mairie de Paris) and by a visiting professorship `ESPCI-Total'. A.P.F. Atman thanks CNPq and FAPEMIG Brazilian agencies for financial funding, and PMMH/ESPCI for hospitality.
\end{acknowledgements}


\appendix
\section{Orthotropic elastic response}
\label{orthotropic}

In this Appendix, we detail elastic calculations on a 2D orthotropic slab of finite thickness $h$. Following the notations of Fig.~\ref{SetUp}, we note $(1,2)$ the orthotropic directions, while $(n,t)$ are the directions respectively normal and tangential to the slab. We note $\tau$ the angle between axes $(1,2)$ and $(n,t)$. For the sake of the computation of the stress profiles in response to a force $\vec F_0$ applied at the free surface, one can switch off gravity, and the mechanical equilibrium of the system writes
\begin{equation}
\partial_n \sigma_{nn} + \partial_t \sigma_{tn} = 0
\qquad \mbox{and} \qquad
\partial_n \sigma_{tn} + \partial_t \sigma_{tt} = 0,
\label{Equil}
\end{equation}
where $\sigma_{ij}$ is the stress tensor. We define the strain tensor $u_{ij}$ from the displacement field $u_i$ as $u_{ij} = \frac{1}{2} (\partial_i u_j + \partial_j u_i)$. It verifies the compatibility condition:
\begin{equation}
\partial_n^2 u_{nn} + \partial_t^2 u_{tt} - 2 \partial_n \partial_t u_{tn} = 0.
\label{Compatibility}
\end{equation}
Introducing the two Young moduli $E_1$ and $E_2<E_1$, the shear modulus $G$ and two Poisson coefficients $\nu_{12}$ and $\nu_{21}$, the generalised Hooke's law relating strain and stress tensors writes, in the orthotropic axes, as follows:
\begin{equation}
\left (
\begin{tabular}{c}
$u_{11}$\\
$u_{22}$\\
$u_{12}$
\end{tabular}
\right )
=
\left (
\begin{tabular}{ccc}
$\frac{1}{E_1}$			& $-\frac{\nu_{21}}{E_2}$		& $0$\\
$-\frac{\nu_{12}}{E_1}$	& $\frac{1}{E_2}$			& $0$\\
$0$					& $0$					& $\frac{1}{2G}$
\end{tabular}
\right )
\left (
\begin{tabular}{c}
$\sigma_{11}$\\
$\sigma_{22}$\\
$\sigma_{12}$
\end{tabular}
\right ).
\label{Wdagger}
\end{equation}
We call $\mathcal{W}_\dagger$ this $3 \times 3$ compliance matrix. It must be symmetric and these coefficients thus verify $\nu_{12}/E_1 = \nu_{21}/E_2$. Elastic energy is well defined if all moduli $E_1,E_2,G$ are positive and $1-\nu_{12}\nu_{21}>0$. In $(n,t)$ axes, we have
\begin{equation}
\left (
\begin{tabular}{c}
$u_{nn}$\\
$u_{tt}$\\
$u_{tn}$
\end{tabular}
\right )
=
\mathcal{W}_\tau
\left (
\begin{tabular}{c}
$\sigma_{nn}$\\
$\sigma_{tt}$\\
$\sigma_{tn}$
\end{tabular}
\right )
\quad \mbox{with} \quad
\mathcal{W}_\tau = \mathcal{Q}^{-1} \mathcal{W}_\dagger \mathcal{Q}
\label{Wtau}
\end{equation}
and the rotation matrix
\begin{equation}
\mathcal{Q} =
\left (
\begin{tabular}{ccc}
$\cos^2\tau$			& $\sin^2\tau$			& $2\cos\tau\sin\tau$\\
$\sin^2\tau$			& $\cos^2\tau$			& $-2\cos\tau\sin\tau$\\
$-\cos\tau\sin\tau$		& $\cos\tau\sin\tau$		& $\cos^2\tau-\sin^2\tau$
\end{tabular}
\right ).
\label{defQ}
\end{equation}
The matrix $\mathcal{W}_\tau$ can be made explicit as follows:
\begin{equation}
\mathcal{W}_\tau = \frac{1}{E_2}
\left (
\begin{tabular}{ccc}
$A$		& $-C$	& $2D$\\
$-C$		& $B$	& $2F$\\
$D$		& $F$	& $H$
\end{tabular}
\right ),
\label{Wtauexplicit}
\end{equation}
with
\begin{eqnarray}
A & = & T \cos^4\tau + \sin^4\tau + 2R\cos^2\tau\sin^2\tau,
\label{defA}\\
B & = & \cos^4\tau + T \sin^4\tau + 2R\cos^2\tau\sin^2\tau,
\label{defB}\\
C & = & \nu_{21} + \cos^2\tau\sin^2\tau (2R-1-T),
\label{defC}\\
D & = & \cos\tau\sin\tau \left[ (\sin^2\tau - \cos^2\tau)R + \cos^2\tau (1+T) - 1 \right],
\label{defD}\\
F & = & \cos\tau\sin\tau \left[ (\cos^2\tau - \sin^2\tau)R + \sin^2\tau (1+T) - 1 \right],
\label{defF}\\
H & = & \nu_{21} - 2\cos^2\tau\sin^2\tau (2R-1-T) + R,
\label{defH}
\end{eqnarray}
and where we have introduced the two dimensionless numbers
\begin{equation}
T = \frac{E_2}{E_1} = \frac{\nu_{21}}{\nu_{12}},
\quad \mbox{and} \quad
R = \frac{1}{2} \, E_2 \left( \frac{1}{G} - \frac{\nu_{12}}{E_1} - \frac{\nu_{21}}{E_2} \right).
\label{defTandR}
\end{equation}

With the four roots $X_k$ ($k=1, ..., 4$) of the biquadratic equation $X^4+2RX^2+T=0$, that is
\begin{equation}
X = \pm \sqrt{-R \pm (R^2-T)^{1/2}},
\label{defXk}
\end{equation}
the general solution of the problem can be written as sums of Fourier modes:
\begin{eqnarray}
\sigma_{nn} (n,t) & = & \sum_{k=1}^4 \int_{-\infty}^{+\infty} \!\!\!\! b_k(q) \, e^{iqt+iY_k qn} dq,
\label{SGsnn}\\
\sigma_{tt} (n,t) & = & \sum_{k=1}^4 \int_{-\infty}^{+\infty} \!\!\!\! b_k(q) \, Y_k^2 \, e^{iqt+iY_k qn} dq,
\label{SGstt}\\
\sigma_{tn} (n,t) & = & - \sum_{k=1}^4 \int_{-\infty}^{+\infty} \!\!\!\! b_k(q) \, Y_k \, e^{iqt+iY_k qn} dq,
\label{SGstn}
\end{eqnarray}
where $Y_k=(X_k-\tan\tau)/(1+X_k\tan\tau)$. The four functions $b_k$ are determined by the boundary conditions at the top and the bottom of the slab.

At the free surface ($n=0$), the overload force imposes two components of the stress:
\begin{equation}
\sigma_{nn} = F_0 \cos\theta \, \Delta(t)
\quad \mbox{and} \quad
\sigma_{tn} = F_0 \sin\theta \, \Delta(t), 
\label{TopBC1}
\end{equation}
where $\theta$ is the angle between $\vec F_0$ and the direction of the $n$ axis (see Fig.~\ref{SetUp}), and where $\Delta(t)$ is a normalised function which tells how this force is distributed along the surface -- e.g. a Dirac or a Gaussian of width $w_F$. We need here its Fourier transform $s(q)$. For the Gaussian case, $s(q) = \frac{1}{2\pi} \exp(-\frac{1}{2} w_F^2 q^2)$. We typically take $w_F \to 0$ (a $\delta$-peak). These top conditions (\ref{TopBC1}) then give
\begin{equation}
\sum_{k=1}^4 b_k = F_0 \cos\theta \, s(q)
\quad \mbox{and} \quad
\sum_{k=1}^4 b_k Y_k = - F_0 \sin\theta \, s(q).
\label{TopBC2}
\end{equation}

At the bottom of the slab ($n=h$), we impose rigid and rough conditions, i.e. vanishing displacements in both $t$ and $n$ directions: $u_t=u_n=0$. In order to get equations on the functions $b_k$, we must transform these conditions into equations on the stress components. Taking its derivative along $t$, the condition $u_t=0$ gives $u_{tt}=0$, i.e.
\begin{equation}
-C \sigma_{nn} + B \sigma_{tt} + 2F \sigma_{tn} = 0,
\label{BotBC1t}
\end{equation}
leading to
\begin{equation}
\sum_{k=1}^4 b_k \left [ -C - 2F Y_k + B Y_k^2 \right ] e^{iY_k q h}= 0.
\label{BotBC2t}
\end{equation}
Similarly, the condition $u_n=0$ gives, after a double derivative along $t$, the relation $2 \partial_t u_{tn} = \partial_n u_{tt}$, leading to
\begin{equation}
\sum_{k=1}^4 b_k \left [ 2D + (C-2H) Y_k + 4F Y_k^2 - B Y_k^3 \right ] e^{iY_k q h}= 0.
\label{BotBC2n}
\end{equation}

The four linear equations (\ref{TopBC2}, \ref{BotBC2t}, \ref{BotBC2n}) can be inverted, leading to large but analytic expressions for the functions $b_k$. Integrations over $q$ involved in Eqs.~\ref{SGsnn}-\ref{SGstn} must, however, be computed numerically. Finally, the stress components, made dimensionless by $F_0/h$, can be plotted for given values of the five parameters $\tau$, $T$, $R$, $\nu_{21}$ and $\theta$, as functions of $t/h$ at a given depth (e.g. $n=h$). We checked that the results are insensitive to the value of $w_F/h$, as long as it remains small.

\end{document}